\documentclass[iicol, sn-mathphys-num]{sn-jnl}
\usepackage[utf8]{inputenc}

\usepackage{hyperref}
\usepackage{graphicx}%
\usepackage{multirow}%
\usepackage{amsmath,amssymb,amsfonts}%
\usepackage{amsthm}%
\usepackage{mathrsfs}%
\usepackage[title]{appendix}%
\usepackage{xcolor}%
\usepackage{textcomp}%
\usepackage{manyfoot}%
\usepackage{booktabs}%
\usepackage{svg}%
\usepackage{algorithm}%
\usepackage{algorithmicx}%
\usepackage{algpseudocode}%
\usepackage{listings}%
\usepackage[hybrid, pipeTables, smartEllipses]{markdown}
\usepackage{tabularx}
\usepackage{booktabs}
\usepackage{array}
\usepackage{mdframed}
\usepackage{enumitem}

\usepackage{graphicx}   
\usepackage{upgreek}    

\newenvironment{readme}
{%
  \begin{mdframed}[
    topline=false, bottomline=false, rightline=false,
    linewidth=2pt, linecolor=gray!60,
    innerleftmargin=8pt, innerrightmargin=8pt,
    innertopmargin=4pt, innerbottommargin=4pt,
    skipabove=6pt, skipbelow=6pt,
  ]%
  \footnotesize
  \linespread{0.95}\selectfont
  \setlength{\parskip}{2pt}%
  \setlist[itemize]{noitemsep,topsep=2pt,parsep=0pt,leftmargin=1.2em}%
  \setlist[enumerate]{noitemsep,topsep=2pt,parsep=0pt,leftmargin=1.5em}%
}{%
  \end{mdframed}%
}

\theoremstyle{thmstyleone}%
\theoremstyle{thmstyletwo}%

\theoremstyle{thmstylethree}%

\usepackage{listings}
\usepackage[most]{tcolorbox}

\begin{document}

\title[Article Title]{Agentic-J: An AI Agent for Biological Microscopy Image Analysis}


\author[1]{\fnm{Lukas} \sur{Johanns}}
\equalcont{These authors contributed equally to this work.}

\author[2]{\fnm{Marilin} \sur{Moor}}
\equalcont{These authors contributed equally to this work.}

\author[1]{\fnm{Davide} \sur{Panzeri}}

\author[1,3]{\fnm{Yu} \sur{Zhou}}

\author[4]{\fnm{Xinyi} \sur{Chen}}

\author[1]{\fnm{Nora F. K.} \sur{Pauly}}

\author[5]{\fnm{Zixuan} \sur{Pan}}

\author[1,6]{\fnm{Matthias} \sur{Gunzer}}

\author[7]{\fnm{Andreas} \sur{Müller}}

\author[5]{\fnm{Yiyu} \sur{Shi}}

\author[2]{\fnm{Hedi} \sur{Peterson}}

\author*[1]{\fnm{Jianxu} \sur{Chen}}\email{jianxu.chen@isas.de}

\affil*[1]{\orgname{Leibniz-Institut für Analytische Wissenschaften – ISAS – e.V.}, \orgaddress{\city{Dortmund}, \country{Germany}}}

\affil[2]{\orgdiv{Institute of Computer Science}, \orgname{University of Tartu}, \orgaddress{\city{Tartu}, \country{Estonia}}}

\affil[3]{\orgdiv{Faculty of Computer Science}, \orgname{Ruhr University Bochum}, \orgaddress{\city{Bochum}, \country{Germany}}}

\affil[4]{\orgdiv{Informatics Institute}, \orgname{University of Amsterdam}, \orgaddress{\city{Amsterdam}, \country{Netherlands}}}

\affil[5]{\orgdiv{Department of Computer Science and Engineering}, \orgname{University of Notre Dame}, \orgaddress{\city{Notre Dame}, \country{United States}}}

\affil[6]{\orgdiv{Institute for Experimental Immunology and Imaging}, \orgname{University of Duisburg-Essen}, \orgaddress{\city{Essen}, \country{Germany}}}

\affil[7]{\orgdiv{Institute of Molecular and Clinical Immunology}, \orgname{Faculty of Medicine}, \orgaddress{\city{Magdeburg}, \country{Germany}}}


\abstract{Biological image analysis increasingly demands integration across heterogeneous tools, programming environments, and domain knowledge that few researchers can command simultaneously. We present Agentic-J, a containerised, multi-agent AI assistant, primarily for ImageJ/Fiji that enables biologists to specify analysis tasks in natural language, from nuclei segmentation and cell tracking to multi-condition quantification. The agent generates executable scripts organised into a documented project structure, so every analysis decision is traceable and the workflow can be reproduced or shared. The specialised sub-agents handle plugin management, code generation, debugging, quality assurance, and statistical reporting. In this paper we introduce the system's design, demonstrate real biological microscopy image analysis workflows, and detailed the technical implementation. The Agentic-J project is available at \url{https://mmv-lab.github.io/Agentic-J/}.
}


\maketitle

\onecolumn
\section{Introduction}\label{sec1}

Modern biology is increasingly an data-driven science. Advances in microscopy now generate datasets much larger than can be inspected or quantified manually. So, it is critical for microscopy data analysis tools for the automated processing of images to be reliable and reproducible. Fiji \cite{schindelin2012fiji} and ImageJ \cite{ImageJ} are among the most established platforms for this task, supported by a large community and a wide plugin ecosystem that covers segmentation, tracking, registration, and much more. Yet this breadth is also a source of friction: plugins often depend on incompatible software versions, their parameters are rarely documented for non-specialists, and assembling individual tools, together with the required plotting and statistics, into a reproducible workflow is difficult for new users. A substantial share of posts on community forums such as image.sc concern environment and plugin-compatibility issues rather than the biology itself \cite{Ruedenetal2019}. Moreover, Fiji is built primarily as a GUI application: its macro language and scripting interfaces exist, but few non-developers learn them, so most users default to manual point-and-click analysis that is slow and hard to reproduce.

The reasoning and planning capability of Large language models (LLMs) can lower this barrier by turning a natural-language description of an analysis goal into an executable and reproducible pipeline. However, general-purpose models have gaps in domain-specific knowledge such as biomedical image analysis, leading to poor tool choices and incorrect parameters \cite{bioimg_benchmark}. Recent benchmarks designed for microscopy-based scientific reasoning show that even frontier multimodal LLMs achieve only around 53\% accuracy on tasks requiring expert image understanding, hypothesis generation, and experiment proposal \cite{microvqa}. Additionally, an agent with unrestricted access to a user's machine may accidentally delete files, overwrite data, or expose private information, often through long generated scripts whose effects a non-expert cannot easily audit \cite{agentsecurity}. Several agentic systems have already addressed parts of this problem in the imaging domain, such as ImageJ Pipeline Builder~\cite{ijpb}, napari-MCP~\cite{napari-mcp}, BioImage.IO Chatbot~\cite{bioimageio}, and general biomedical agents such as Biomni~\cite{biomni_paper}. However, none of these systems combine sandboxed execution, managed plugin dependencies, and curated domain grounding specially for bioimage analysis in a single deployable system.

In this work, we introduce Agentic-J, a containerized multi-agent system for reproducible
biological image analysis in Fiji. A newcomer can run a complete analysis without
managing software, choosing between guided UI instructions and automated script
generation, while the agent operates inside a sandbox and records each project as
a self-contained folder of data, scripts, and step-by-step documentation. The open-source code is available at \url{https://github.com/MMV-Lab/Agentic-J}. In general, the main features of Agentic-J are highlighted below:
\begin{itemize}
    \item an accessible single-package system permits biologists to run complete bioimage analysis workflows in Fiji through natural language instructions, without installing or configuring any additional software;
    \item a human-in-the-loop design that keeps the familiar Fiji interface, a conversational agent, and a reproducible project record in one place, so that bioimage analysts stay in control;
    \item safe execution by sandboxing the agent and bundling a curated set of compatible plugins, so analyses run reliably and cannot affect the user's base operating system;
    \item domain-aware reasoning that grounds the agents in curated bioimage analysis knowledge and improves over time by reusing past successes and avoiding repeating failure;
    \item standards compliance, with each project automatically audited
    against community standards for publishing images and analyses.
\end{itemize}

The remainder of this paper is organised as follows. Section~2 describes the
capabilities of Agentic-J and demonstrates them on real-world biological problems, including its reproducibility, quality-assurance, and teaching-oriented features. Section~3 then turns to the technical setup, covering the containerization, the multi-agent system, and the knowledge bases. Finally, Section~4 discusses the current limitations and outlines future works.

\section{Agentic-J in practice}

\subsection{Usability of Agentic-J}

\begin{figure*}[bt]
    \centering
    \includegraphics[width=\linewidth]{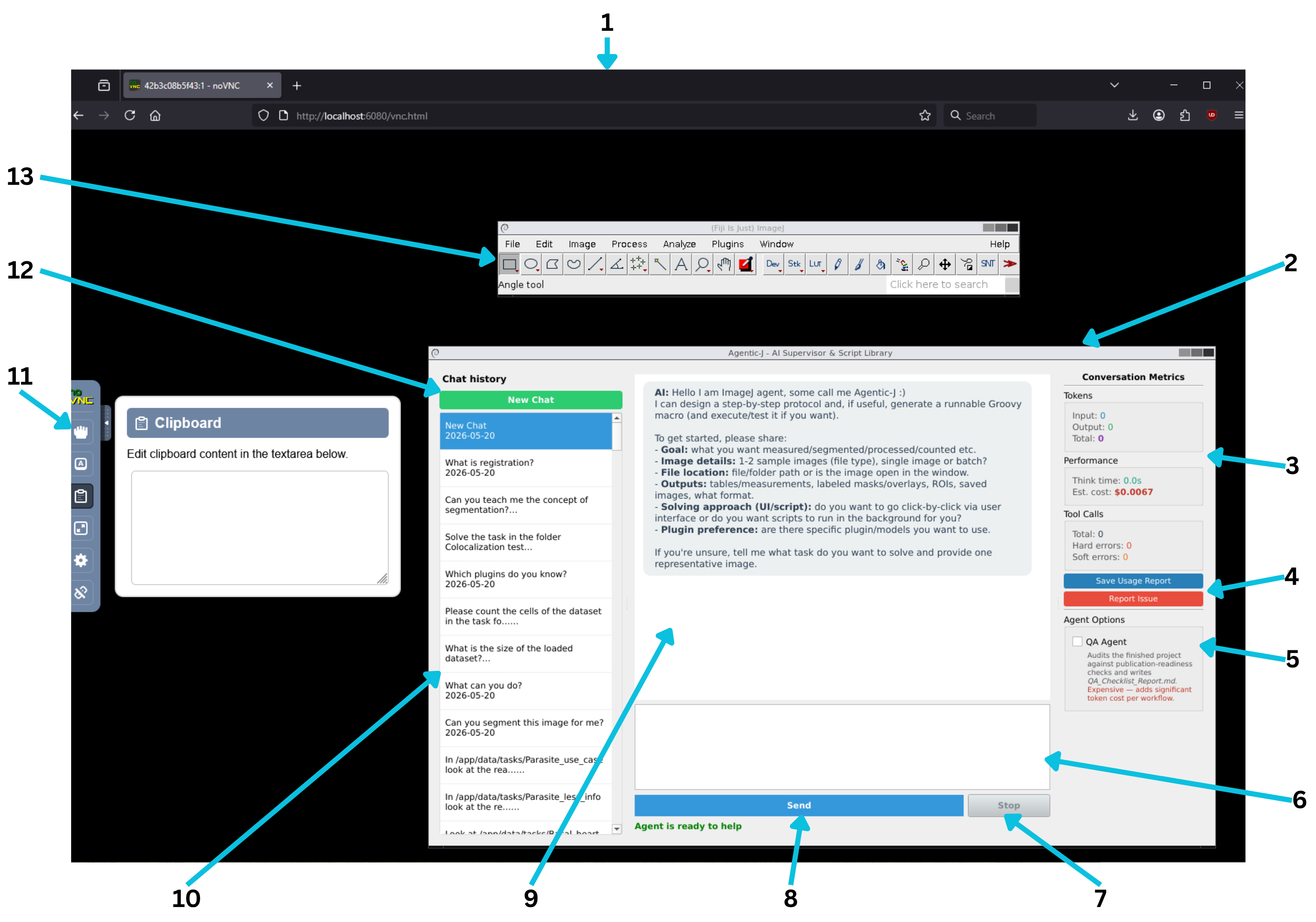}
    \caption{The Agentic-J interface from a browser view (1). In the middle, the Fiji task bar is available and can be accessed directly (13). Below is the chat interface, which the user uses to interact with the agent (2). The interface also includes previous chats that users can revisit and continue for topic-related discussions (10) and the option to start a new chat (12). On the right side of the interface, users can track the spent tokens, estimated cost, and agent runtime for the current chat thread (3). It is also possible to send an error report to the developers or save the project for reproducibility (4). Additionally, the QA agent can be activated to create a community-standard quality assessment of the project (5). In the message field (6) the user can ask the agent a question, by pressing the send button (8) and will receive an answer in the chat window (9). The agent can be interrupted with the stop button (7). The appearance and resolution of the window is controlled through the sidebar (11), where the user also can copy the clipboard from their computer.
    }
    \label{fig:chat_ui}
\end{figure*}

One of the defining features of Agentic-J is that it brings together tools that would ordinarily require significant technical effort to install and maintain and makes them available as a single, self-contained system, accessible via natural language instructions. As illustrated in Figure \ref{fig:chat_ui}, the end user interacts with a familiar chat interface in the browser, while the full Fiji workspace is visible in the same tab; the technical integration happens invisibly in the background.

A key usability challenge in bioimage analysis is that tools developed by different research groups often conflict with one another: deep learning segmentation models such as Cellpose and StarDist require different software versions that cannot coexist in a single installation. Agentic-J resolves this by packaging each tool in its own isolated environment, baked into the system at build time, so that switching between segmentation approaches requires no manual intervention from the user. The Fiji image analysis platform runs inside the same system through a direct software bridge, which mean the agent can open images, run plugins, and collect measurements without the user having to manage a separate application. The agent itself operates as a coordinated team of specialists subagents: a main supervisor agent understands the scientific goal delegates tasks to dedicated agents for code generation, error correction, and statistical analysis, etc. A built-in knowledge base populated with plugin documentation, scripting guides, and a record of previously encountered errors and their solutions, allows the agent to learn from accumulated experience rather than starting from scratch on each task. The system is compatible with both OpenAI and a range of other model providers \cite{openai_api, openrouter_api}, so it is not tied to a single model vendor.

The system runs equally on Windows, Linux, and macOS without any platform-specific configuration. Users provide their image data through a dedicated data folder, and all outputs, processed images, analysis scripts, result tables, and figures, are collected in the same location, making it straightforward to trace what was done and retrieve results for downstream use. Also, chat history is preserved across sessions. The user can return to any previous project, review what analysis was performed, and continue from where they left off, with all associated scripts, figures, and data still linked to that conversation thread.

The architecture is also designed with future extensibility in mind. As biological questions increasingly require tools beyond ImageJ, the system can communicate with external platforms through standardised protocols. Rather than requiring a dedicated integration for each new tool, any external platform that supports the Model Context Protocol (MCP) can expose its functionality to the Agentic-J supervisor, which then treats those capabilities as its own. For example, three-dimensional visualisation in napari \cite{napari}, is implemented through napari-MCP \cite{napari-mcp}. Napari is the first such use case, but the mechanism is not specific to any single software.

\subsection{Image Analysis Workflows for Real-world Biological Problems with Agentic-J}

 To evaluate the practical utility of Agentic-J in real-world scenarios, we tested the system on two distinct workflows derived from peer-reviewed biological studies. These case studies are selected to demonstrate the two primary operational modes of the system. The first workflow illustrates how the agent provides step-by-step UI-guided instructions to help users navigate complex graphical plugins. While the second example shows the agent's capacity to autonomously generate, execute, and document multi-step analysis scripts. Together, these examples provide an initial assessment of how the system can assist researchers in translating biological questions into reproducible image analysis pipelines.

\subsubsection{Tracking mouse stem cells (reproducing a workflow from a paper using UI)}

Celltraxx is an automatic workflow for tracking cell migration in phase contrast images~\cite{Holme2023}, where the performance was compared with manual processing as well as TrackMate~\cite{Ershov2021} using different datasets and tasks. One example was to analyze an image series of mouse muscle stem cells in hydrogel microwells. Provided the parameters and methods in the paper, we attempted to reproduce the analysis in Agentic-J (using TrackMate, via UI steps) to verify if one can use the same configuration to replicate the results. Specifically, we set up the experiment with the last 100 images from the aforementioned mouse muscle stem cell series in the data/mouse\_cell\_tracking folder, while the parameters/methods were collected from the paper and described as one prompt:
\begin{tcolorbox}[
    title=Prompt,
    colback=gray!10,
    colframe=black,
    sharp corners,
    boxrule=0.5pt
]

I have a list of images in the \texttt{mouse\_cell\_tracking} folder and I would need to find the number of tracks using TrackMate. These 16-bit images should first be converted to a binary image and then inverted to get white spots (255) on a grey background (123) such that the TrackMate thresholding detector could be used with a threshold grey level of 150 to easily distinguish all 2000 spots in this image series. The LAP tracker should be used for tracking the spots, but with more liberal gap closing settings in order to track the fast-moving spots as well as possible:
\begin{itemize}
    \item max frame gap = 4
    \item alternative linking cost factor = 1.05
    \item linking max distance = 60 pixels
    \item gap closing max distance = 100 pixels
    \item allow gap closing = true
    \item allow track splitting = false
    \item allow track merging = false
\end{itemize}

Could you give steps using UI?

\end{tcolorbox}

After initial instructions, the user prompted extra information, as the plugin interface did not have all of the Linear Assignment Problem (LAP) tracker parameters that were initially required (likely due to a difference in TrackMate versions). The feedback from the agent helped to achieve the same results as the ones reported by the original authors \cite{Holme2023}: the same number of tracks (29) and spots/cells (2001).

\begin{figure*}[htbp]
    \centering

 \begin{minipage}[t]{0.49\textwidth}
    \centering
    \includegraphics[height=5cm]{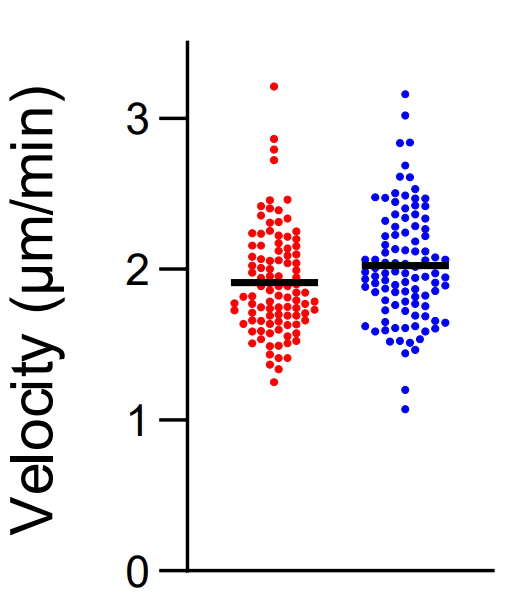}
\end{minipage}
\begin{minipage}[t]{0.49\textwidth}
    \centering
    \includegraphics[height=5cm]{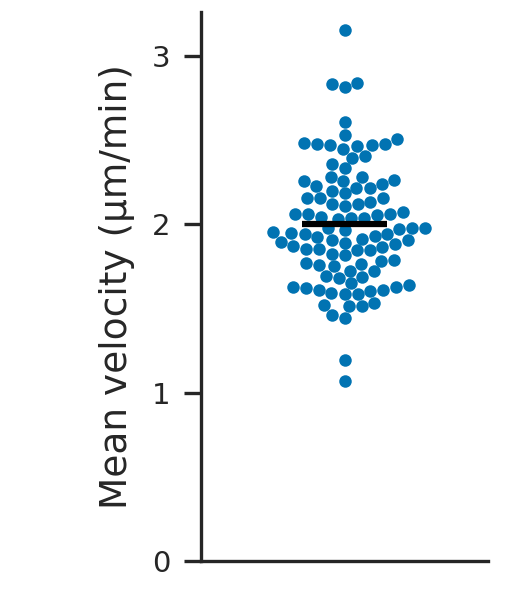}
\end{minipage}

\caption{On the left, the TrackMate (points in blue) distribution achieved in the peer-reviewed publication\cite{Holme2023}. This figure is reused under CC BY 4.0. On the right, the TrackMate distribution generated via Agentic-J can be seen on the right. Each dot represents the mean velocity from one time point to the next. The black line shows the overall mean velocity.}
\label{fig:swarm_plots}

\end{figure*}

The paper also reported the mean velocities ($\mu$m/min) from image to image as a (bee)swarm plot (Figure S3e in \cite{Holme2023}). In order to verify the capability of plotting in Agentic-J, the subtitle of the supplementary figure was adapted with physical unit details (initially only pixels information) and the type of plot with initial design features was described in the following prompt. 
\begin{tcolorbox}[
    title=Prompt,
    colback=gray!10,
    colframe=black,
    sharp corners,
    boxrule=0.5pt
]

I added the tracks csv to the initial data folder. Could you help me do a swarmplot with the comparison of the mean velocities from image to image calculated from tracking by TrackMate analysis of the corresponding GT images (dots in blue).
Each dot represents the mean velocity from one time point to the next. The black line shows the overall mean velocity. I found out that the physical unit is 0.645 um per px and the time is 5 minutes between images.

\end{tcolorbox}
The first generated plot was modified using the chat interface by asking it to edit the plot style, i.e. removing x-axis labels, re-setting the y-axis to start from 0, ensuring that the increment is 1 etc. It is important to note that with the GPT-5.2-based plotting agent, the plotting style should be prompted as clearly as possible and repeated if necessary. Although some elements can be kept constant, successive interactions may still produce variations in the code or introduce changes to other sections. As can be seen from Figure \ref{fig:swarm_plots}, the imitation was quite successful, while the jitter style differs, the dots are following a similar distribution and the horizontal black line is situated in the same location. For the complete task (tracking task + plotting (to a satisfactory level)), the estimated thinking time was 7 minutes, this included iterative re-designing done by the plotting agent. On a side note, for the sake of clear visualisation in this paper, the generated script provided by Agentic-J was modified further (font-size of the axis texts and making the plot relative to the output size), and run independent of the system. Full prompts, generated scripts, and workflow outputs are included in the Appendix \ref{supp_mat}.

\subsubsection{Apoptotic behaviour of parasite-infested cells (building a multi-step workflow)}

To demonstrate the coding and multi-step workflow capabilities of Agentic-J on a real-world microscopy image analysis task, the system was evaluated on a task derived from a peer-reviewed publication~\cite{parasite_paper}. The biological objective of the original study was to investigate whether intracellular proliferation of the \textit{Leishmania major} parasite is associated with apoptotic behaviour of the infected host cell. We used the original image data and ROI annotations, provided by the authors of the study, for testing if Agentic-J is able to automatically build multi-step workflows. 

\paragraph{Provided data}

The dataset comprised two acquisitions of the same field of view, summarised in Table~\ref{tab:parasite_data}. Besides the raw images, to test the exact scenario in the original study, where biology experts manually annotated parasite in the image as ROI, Agentic-J received manually annotated parasite ROIs, while the determination of apoptotic host-cell status was left entirely to the generated workflow. Automatic parasite detection is also possible in theory, but may require additional training of an object detection model. Arguably, in practice, manual ROI selection or annotation is still much faster to get the results than building new models, and therefore still commonly used by biologists.

\begin{table*}[!hb]
\centering
\caption{Channel composition of the two acquisitions provided for the parasite
workflow. The live-cell movie was acquired over multiple time points at a
temporal resolution of 10 minutes per frame; the fixed-endpoint image captured
the same field of view after fixation and TUNEL staining.}
\label{tab:parasite_data}
\begin{tabular}{@{}llll@{}}
\toprule
Acquisition & Channel 1 & Channel 2 & Channel 3 \\
\midrule
Live-cell time-lapse & Phase contrast & mKikume green & mKikume red \\
 & (host cells) & (parasites) & (parasites) \\
\addlinespace
Fixed-endpoint TUNEL & Phase contrast & TUNEL signal & DAPI \\
 & (host cells) & (apoptotic nuclei) & (all nuclei) \\
\bottomrule
\end{tabular}
\end{table*}

\paragraph{Workflow objectives}

The agent was asked to carry out the analysis in four connected steps. It first had to align the final phase-contrast frame of the live-cell movie with the phase-contrast channel of the fixed TUNEL image and apply the same transformation to all remaining channels, bringing the two acquisitions into a common coordinate frame. Working on these aligned images, it then measured fluorescence intensities from the provided parasite and background ROIs to calculate the proliferation index. Each parasite then had to be linked to its host cell and classified as TUNEL-positive or TUNEL-negative. Finally, drawing all of this together, the agent had to generate statistical analyses and publication-style plots evaluating the relationship between parasite proliferation and apoptotic host-cell status. For each parasite ROI, the proliferation index was defined as:

\[
10 - \frac{R}{G}
\]

where \(R/G\) corresponds to the ratio between the background-corrected red and green fluorescence intensities measured within the parasite ROI.

\begin{figure*}[htpb]
    \centering
    \includegraphics[width=\linewidth]{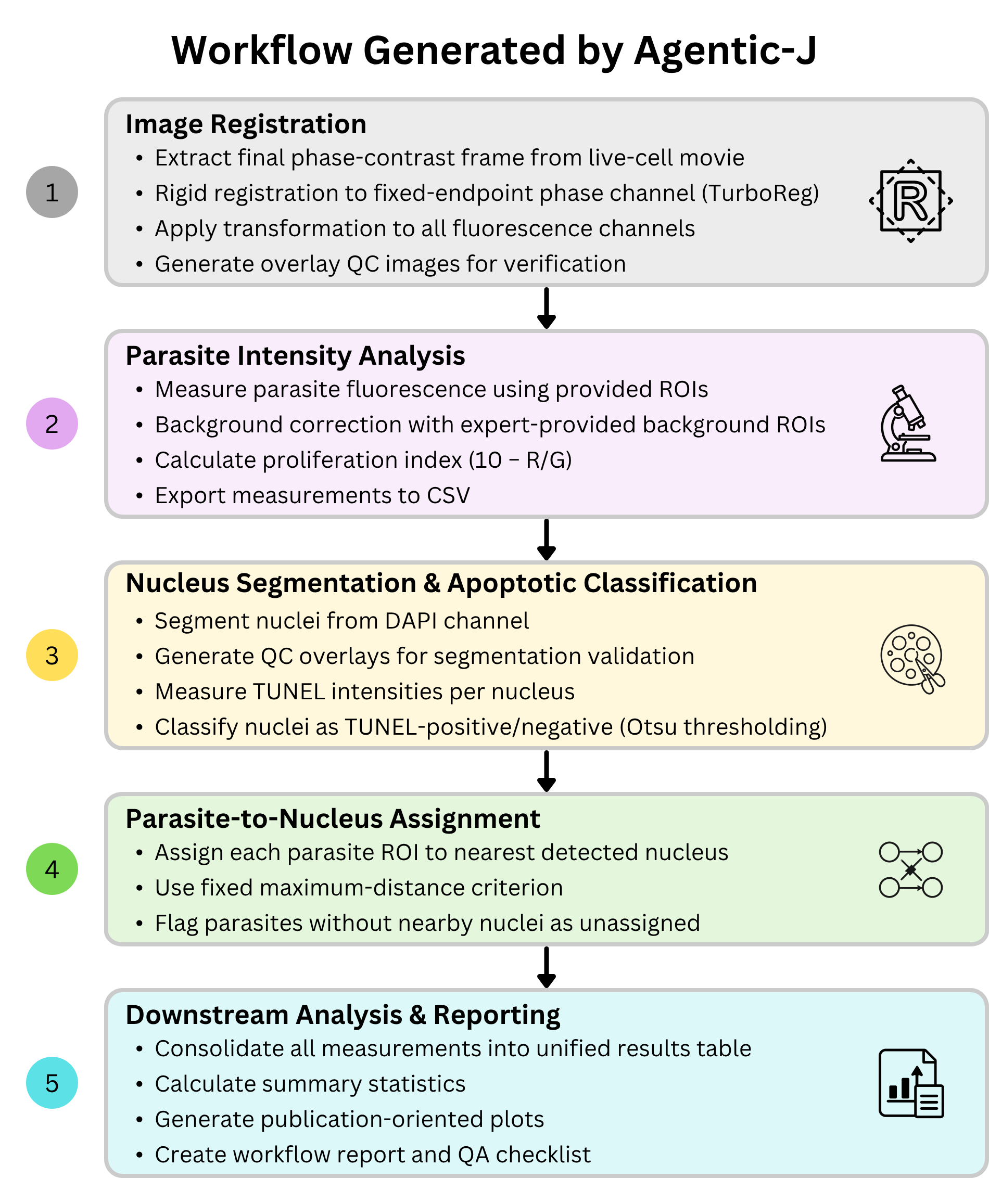}
    \caption{The end-to-end analysis workflow autonomously generated by Agentic-J for
the parasite apoptosis task, comprising five sequential stages, each reported
automatically to the user: (1) rigid registration of the fixed-endpoint image to
the live-cell movie via TurboReg \cite{turboreg}, with the transform propagated to all channels;
(2) ROI-based, background-corrected measurement of parasite fluorescence and
calculation of the proliferation index $10 - R/G$; (3) DAPI-based nucleus
segmentation and Otsu-thresholded TUNEL-positive/negative classification;
(4) assignment of each parasite to its nearest nucleus within a fixed maximum
distance; and (5) consolidation of results, statistics, publication-ready plots,
and automated workflow and QA reports.}
    \label{fig:workflow_parasite}
\end{figure*}

\paragraph{Generated workflow} 

\begin{figure*}[htbp]
    \centering
    \includegraphics[width=\linewidth]{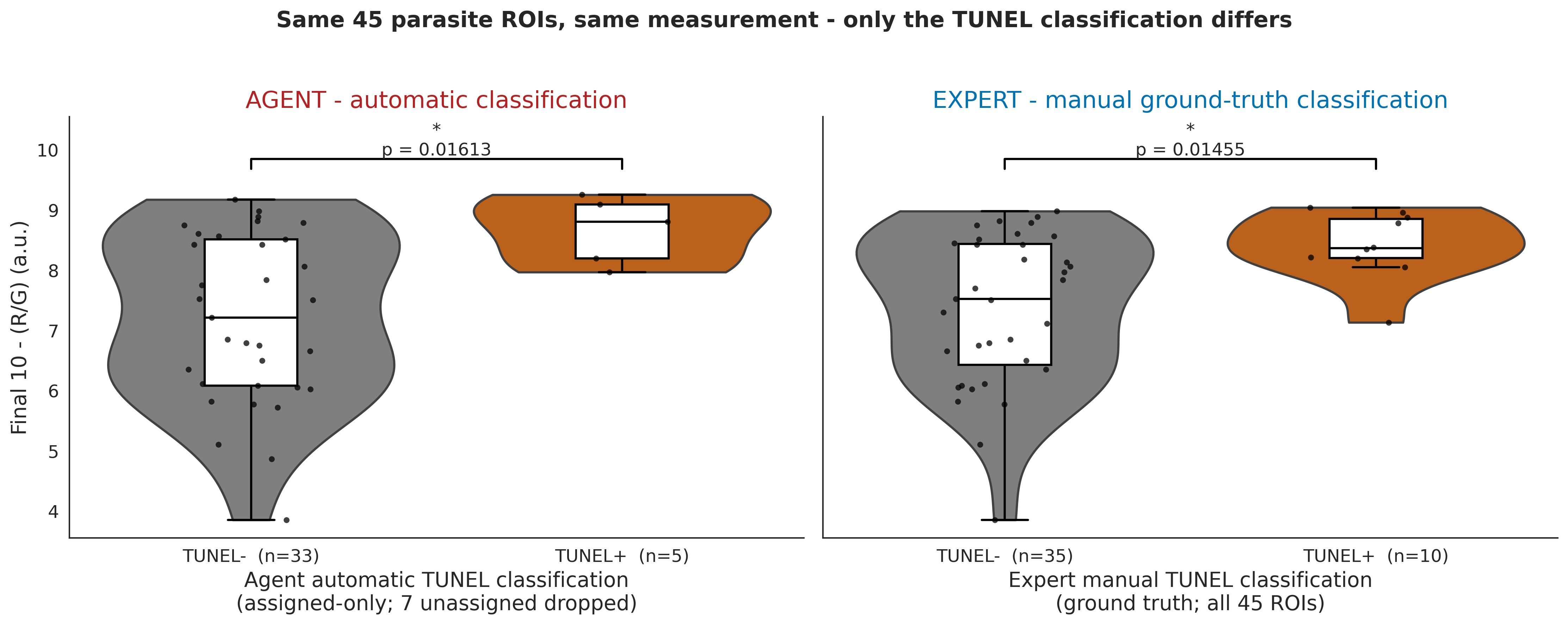}
    \caption{Proliferation index ($10 - R/G$) for the parasite ROIs, grouped by
TUNEL classification. The underlying intensity measurements are identical in
both panels; only the assignment of each parasite to a TUNEL-positive or
TUNEL-negative host cell differs. Left: the agent's automatic classification,
restricted to the 38 parasites that could be assigned to a nucleus (7 unassigned
ROIs dropped). Right: the expert manual ground-truth classification over all 45
ROIs. In both cases the TUNEL-positive and TUNEL-negative populations are
significantly separated ($p = 0.016$ and $p = 0.015$, respectively), showing
that the agent-derived workflow reproduced a statistically similar separation
between the two populations.}
\label{fig:agent_vs_expert}

\end{figure*}

Agentic-J autonomously generated a complete end-to-end analysis workflow that is visualized in Figure \ref{fig:workflow_parasite}. The agent successfully assigned 38 out of the 45 provided parasite ROIs and reproduced a statistically similar relationship between proliferation index and apoptotic status compared to the expert annotations (Figure \ref{fig:agent_vs_expert}). However, approximately 50\% of the TUNEL-positive parasites could not be assigned to a nearby nucleus, resulting in reduced sensitivity (see Figure \ref{fig:classification_agreement}). Overall classification accuracy reached 88.89\%, with a specificity of 100\% and a sensitivity of 50\%.

Importantly, the workflow autonomously identified and documented these shortcomings. The agent proposed several plausible explanations, including an overly restrictive parasite-to-nucleus assignment distance and inaccuracies in TUNEL thresholding. Another reason could be non-rigid morphological deformations introduced during fixation. Although rigid registration was successfully performed, local cellular deformations likely altered the spatial relationship between parasite ROIs and apoptotic nuclei.

A knowledgeable user can inspect the generated quality control (QC) overlays and iteratively refine parameters such as the parasite-to-nucleus assignment distance. This experiment therefore highlights both the strengths and limitations of autonomous microscopy workflows: while the agent was capable of generating a reproducible end-to-end analysis pipeline with biologically meaningful results, expert-guided refinement remained critical for resolving ambiguous edge cases, to deliver trustworthy outputs.

\subsection{Reproducibility and quality assurance}

Reproducible reporting of microscopy pipelines requires that authors disclose not only their results but also the data, software, parameters and figure-preparation choices that produced them. To embed these expectations directly in the agent system, we implement a dedicated quality-assurance (QA) agent that audits each completed project against the community-developed checklists for publishing images and image analyses~\cite{comm_checklist}. Items on the checklist are stratified into three tiers, namely minimal (required for publication), recommended (strongly encouraged) and ideal (best practice), and they cover both the analysis workflow (citation of components and platform, processing sequence, key settings, exact software versions, documentation of manual ROIs, availability of example data and code) and the image publication itself (format and cropping, channel and color choices including grayscale and colorblind-accessible palettes, scale bars and annotation legibility, lossless sharing and deposition in dedicated archives).

The QA agent is invoked once, automatically, at the end of every project, and is strictly read-only. It reviews all project outputs, including scripts, documentation, tables, and figures, and grades each item (pass, partial, or fail) adding a brief explanatory note for each. The audit is consolidated into a single \texttt{QA\_Checklist\_Report.md} placed in the project root, which closes with a prioritized to-do list separating critical failures from recommended improvements and ideal additions. By integrating the checklist directly into the analysis loop, the agent acts as a vehicle through which the community's reproducibility guidelines are disseminated and applied after the pipeline is completed.

The current QA agent can be viewed as a proof-of-concept for community standard dissemination. Further guidelines, e.g., specifically for migration data, or how to report the usage of generative AI, can be further integrated when available.

\subsection{Teaching-oriented design and domain-specific guidance}

One of the goals of the Agentic-J system is to enable end-users to independently reason about and perform image analysis on their own. A prerequisite for reaching this goal is lowering the barrier to entry: research on online education consistently identifies the creation of accessible learning environments as essential to effective learning. The flexibility of pace, time and psychological anonymity also play a supportive role \cite{Kiran2025}. The straightforward chat interface and easy setup of the Agentic-J system directly serve this principle, allowing users to engage with the specifics of bioimage analysis in Fiji without configuration overhead. 

The knowledge base underpinning the system's agentic reasoning is built predominantly on education-oriented open-source materials. This includes the open textbook ``Introduction to Bioimage Analysis'', which aims to explain concepts in image analysis for biologists~\cite{Bankhead2023}. The knowledge base also draws on the community-curated Bio-image Analysis Notebooks, a CC-BY 4.0 collection of Jupyter notebooks covering programmatic image analysis, machine-learning methods, and quantitative image processing~\cite{haase2024bioimage}. This RAG-based grounding in educational materials is shown to have potential in personalized learning. A short-term educational pilot study~\cite{Nmeth2025} indicated that the topics covered in embedded materials in RAG are less prone to hallucinations. This could be linked to another result from the same study, where higher student engagement and motivation  to learn was qualitatively recorded~\cite{Nmeth2025}.

One concrete example of domain-specific guidance in Agentic-J is the plugin parameter explanation feature. When a user gets stuck choosing values in a plugin dialog window, the system uses a separate vision-language model to translate the plugin interface into text, combines this with the plugin's documentation, and considers the current image context to provide targeted recommendations with explanations. Optimal fine-tuning of parameters in image analysis is critical, and expert-informed, human-in-the-loop guidance has been demonstrated to improve results in microscopy workflows \cite{Roels2020}. By providing context-aware parameter advice with explanations, the system supports users in developing a transferable understanding of image processing decisions rather than merely executing steps.

A further avenue for learning is built into the system's reproducibility infrastructure. Because all analysis steps are saved as Groovy/Python scripts, users have the option to inspect, adapt, and reason about the generated code. Code review is recognized as a straightforward and effective technique for accelerating learning and reducing potentially costly errors \cite{Rokem2024}. Users may also submit their own scripts for diagnosis, turning debugging into an active learning experience. 

Finally, where a generalist model might default to broad, high-level answers, Agentic-J is primed specifically for bioimage analysis. Its knowledge base comprises curated educational texts, plugin documentation in the form of human-authored API reference sheets, and ImageJ scripting resources (Groovy). 

\section{Agentic-J technical setup}
\subsection{The overall agentic system}

\begin{figure*}[htbp]
    \centering
    \includegraphics[width=\linewidth]{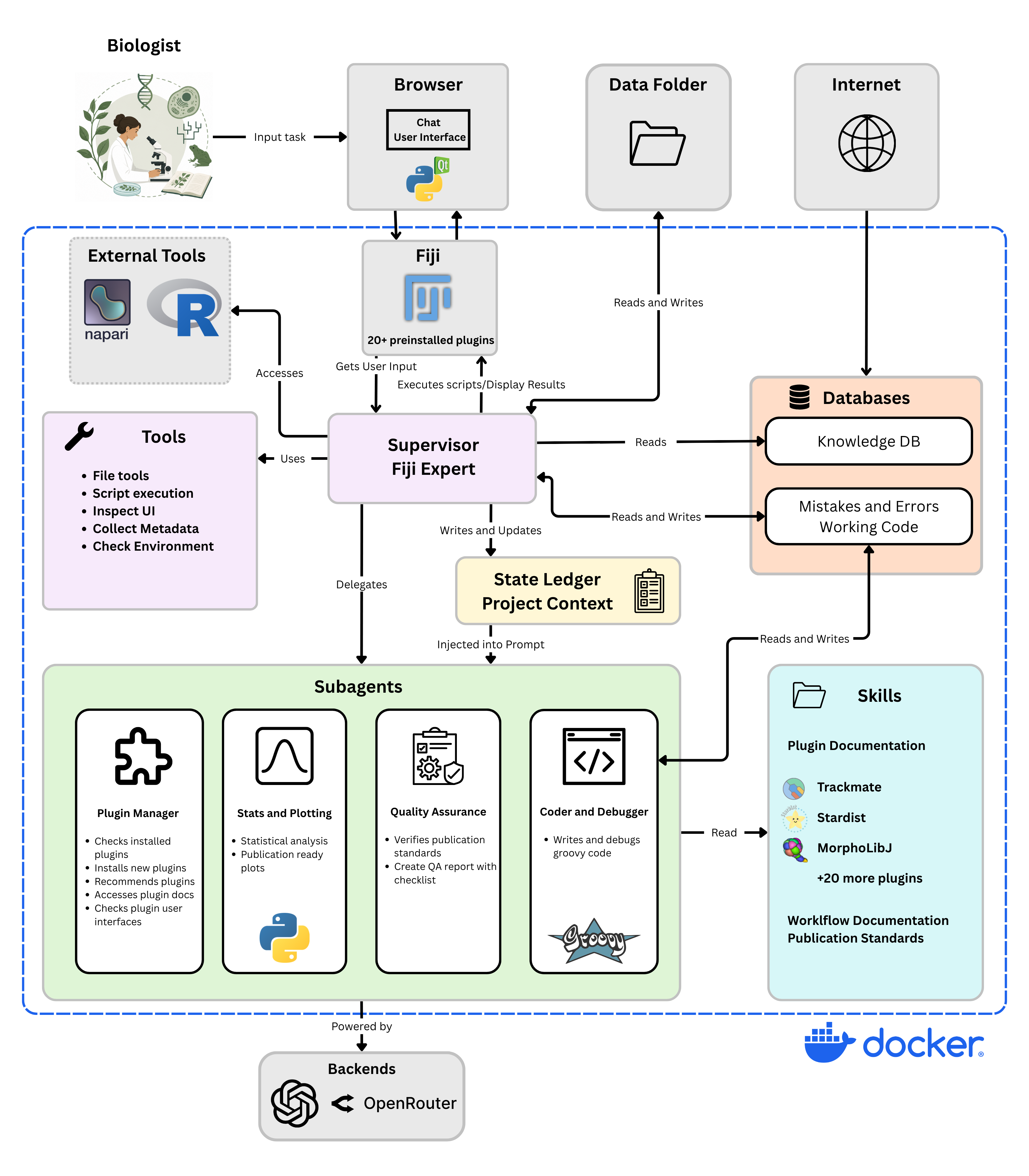}
    \caption{Structure of the Agentic-J system. A biologist submits a task via a chat interface \cite{pyqt} to a supervisor agent, which manages file operations, metadata extraction, and direct Fiji code execution, with
  access to a local data directory and a knowledge database. A state ledger storing pipeline instructions, metadata, and the scientific goal is shared across four subagents: a plugin manager, a coder–debugger, a
   plotting agent, and a quality assurance agent. Subagents draw on a shared skill repository of plugin documentation and workflow guidelines. The system is containerised with Docker and supports
  the OpenAI \cite{openai_api} and OpenRouter \cite{openrouter_api} APIs; external tools such as Napari via MCP \cite{napari-mcp} can be integrated as extensions.  All registered trademarks, brand names, and logos (including Python, Docker, R, Fiji, OpenAI, OpenRouter and others) are the property of their respective owners. Their use     
  here is solely for nominative, descriptive purposes to identify the integrated technologies and does not imply affiliation or endorsement.}
    \label{fig:setup}
\end{figure*}

An overview of the system architecture is provided in Figure \ref{fig:setup}. The system operates via API calls through an OpenRouter or OpenAI-compatible key, and users interact with it through a conversational chat interface.
All incoming requests are first handled by a LangChain deep agent \cite{langchain}, which constructs a structured execution plan called the state ledger. This ledger is shared across all agents to keep the plan transparent and coordinated throughout the pipeline. The supervisor agent has access to a knowledge database to support context-aware planning. To protect user privacy, the agent does not have direct access to raw image data; instead, it retrieves file metadata and invokes dedicated file tools to gather the necessary input context.

Plugin-related queries are routed to a plugin manager agent, which has visibility over both the installed plugin set and a broader plugin registry, enabling it to suggest additional tools beyond the curated container when needed. Once planning is complete and any additional context has been collected from the user, such as the execution mode (script-based or UI-guided) and the intended processing pipeline, the system may invoke one or more specialized agents:

\begin{itemize}
\item The coder agent generates scripts by consulting curated plugin documentation and domain-specific skills.
\item The debugger agent performs iterative script correction.

\item The plugin manager provides plugin-specific guidance, such as selecting the correct parameters in UI dialogs.

\item An optional Python-based data analyst agent supports downstream quantitative analysis.
\item Users may also request a QA agent, which evaluates whether the project meets established publication standards for reproducibility and figure correctness \cite{comm_checklist}.
\end{itemize}

Each subagent is equipped with a targeted toolset: covering plugin UI inspection, file reading, Fiji Log window parsing, and more, thereby enabling reliable, domain-specific assistance at each stage of the workflow.

\subsection{Containerisation and security}

The Fiji ecosystem offers a large and varied collection of plugins, numbering in the hundreds, which, while a significant strength, introduces a non-trivial maintenance burden for end users. Ensuring plugin compatibility is essential to a stable experience, and is one of the primary motivations for providing a containerised environment with a curated set of pre-installed plugins.

Plugin selection was carried out empirically using two complementary approaches: (1) citation-based analysis, examining how frequently each plugin's associated paper was cited, as well as how often the plugin was mentioned in the methods sections of other works; and (2) download statistics from the ImageJ site tracker, which logs the total number of unique update requests per plugin \cite{imagejUpdateSitesJson}. The top 30 non-built-in plugins were identified by aggregating these rankings. The resulting set includes widely-used tools such as MorphoLibJ \cite{Legland2016}, StarDist \cite{Schmidt2018}, TrackMate \cite{Ershov2021}, and Labkit \cite{Arzt2022}.

A recurring theme in the image.sc community forums further reinforces the need for containerisation: a notable proportion of Fiji-related posts concern plugin environment issues, including misconfigured Python environments and incorrect path settings \cite{Ruedenetal2019}. Bundling plugins into a single pre-configured container removes this friction entirely, users do not need to manage dependencies, virtual environments, or plugin paths themselves.

The most pressing motivation for containerisation, however, concerns security. Protecting users' sensitive biological data and constraining unpredictable agent behaviour are both critical considerations. At the time of development, the rise of autonomous agent frameworks similar to OpenClaw \cite{openclaw}, brought with it growing concerns around unintended system access and harmful side effects. Confining the agent's filesystem access to a dedicated \texttt{data/} directory, with all script generation and execution taking place inside the container, substantially reduces this risk. Prior work has highlighted the dangers of running agents with unrestricted local filesystem access \cite{agentsecurity}; of particular concern is the possibility that long generated scripts (e.g. in Groovy) may contain destructive file operations that a non-expert user would be unlikely to catch. While static rule-based guardrails can help restrict agent behaviour, their reliable enforcement is not guaranteed. Research has shown that the pattern matching technique of static rule-based guardrails (e.g. Llama-Guard) can be easily obfuscated \cite{guardrails}.

An additional benefit of the unified plugin environment is improved agentic support: rather than relying on general pre-training knowledge, agents reference curated plugin documentation to produce accurate scripts and correct UI dialog guidance. 

More technically, the container packages a full Fiji/ImageJ distribution with pre-installed plugins, alongside pre-packaged Cellpose model weights and a Conda-managed Python environment providing pyimagej, Qdrant-backed RAG, and LangGraph for agent orchestration. At startup, the entrypoint pre-warms the jgo/Maven dependency cache to eliminate first-run network fetches, thereby bypassing potential SciJava downtime. The whole setup is exposed through a virtual desktop over noVNC on port 6080 running as a non-privileged user with all Linux capabilities dropped.

\subsection{Databases}

\subsubsection{Knowledge database}

General-purpose LLMs are trained on broad web corpora and consequently underperform on domain-specific tasks such as bioimage analysis or the authoring of ImageJ macros. To close this gap, the agent is coupled with a dedicated knowledge base that supports retrieval-augmented generation (RAG)~\cite{RAG_paper}. The base is populated with open-source ImageJ training material, community best-practice guides, biomedical image-processing textbooks, and accompanying Jupyter notebooks; the full inventory is listed in the Appendix \ref{app:rag-inventory}. Documents are chunked at ingestion time and stored in a hybrid Qdrant collection combining sparse and dense embeddings, so that both keyword and semantic queries are supported. At runtime, the agent issues queries through a retrieval tool that fuses the two rankings via Reciprocal Rank Fusion (RRF)~\cite{RRF_paper} and returns the top-scoring chunks. Users may additionally upload PDFs during a session. A naive chunking pass makes the document immediately queryable, while a higher-quality embedding job runs in the background and transparently replaces the provisional entry once it completes.

\subsubsection{Errors and recipes}

To let the agent benefit from prior runs, two complementary stores are maintained, both readable and writable by the relevant agents: a recipe database of working code and an error database of past failures. Whenever a snippet produced by the coder agent executes successfully, the supervisor commits it to the recipe database together with metadata describing the language, the required inputs, and an occurrence count. The debugger writes to both stores in the same fashion. The corrected snippet enters the recipe database, while the originating failure is appended to the error database together with its error type, the offending class, and the broken code. On subsequent runs, the agent can therefore consult both known-good solutions and known failure modes before attempting a new implementation.

\subsubsection{Skills}

Vector databases enable low-latency retrieval but are inherently constrained by chunk size and by the loss of structural context between fragments. To complement RAG with structured, hierarchical knowledge, we adopt the \emph{skills} pattern~\cite{agentskills}. A skill is a directory containing documentation, ready-to-run scripts, and usage instructions, fronted by a \texttt{SKILL.md} file whose header alone is exposed to the agent. From this header the agent decides whether the skill is relevant and, if so, progressively opens the remaining files on demand, lifting the chunk-size ceiling without overloading the context window. This mechanism is used throughout the system. The supervisor's context budget is kept small by outsourcing workflow descriptions and pipeline guidelines to dedicated skill files referenced from its system prompt. The plugin manager and the coder/debugger agents share a collection of plugin-specific skills, each comprising an overview \texttt{SKILL.md}, scripting and UI usage notes, and a sample workflow in Groovy or as a UI recipe. Finally, the statistics and plotting agent draws on a dedicated skill that codifies our conventions for publication-ready figures.

\section{Discussion}

In this work, we presented Agentic-J, a containerised multi-agent system that, in a single deployable package, brings together sandboxed execution, managed plugin dependencies, and curated domain grounding for bioimage analysis. Just as central to the design, however, is what Agentic-J is not: it is not a terminal-only agent that displaces the analyst. Practical microscopy analysis is an inherently visual and iterative activity, thus our system is built around this philosophy. The familiar Fiji GUI, a conversational chat, and an agentic backend share a single browser tab and a single project folder, so a researcher can ask the agent to draft a Groovy pipeline, intervene through a plugin dialog when biological judgement is required, and request a re-plot or a parameter explanation in the same conversation. Analysis in Agentic-J becomes a collaboration between a human expert and an agent rather than an autonomous handover, what we believe matches how microscopy data is actually interpreted in practice, and one that distinguishes the system from fully autonomous pipelines that offer no natural point of intervention. Each project also leaves behind an automated audit against the publication checklists of Schmied et al. \cite{comm_checklist}, to our knowledge the first runtime implementation of those checklists, closing a gap between community-established standards and day-to-day practice so that reproducibility is produced alongside the analysis rather than retrofitted afterwards.




The attributes described in the preceding chapters represent a technical foundation, but their practical value can only be established through structured collaboration with domain experts. Evaluating an agentic bioimage analysis system requires more than benchmark accuracy on curated datasets; it requires understanding whether the system meaningfully reduces the cognitive and technical burden that biologists currently bear when translating experimental questions into analysis pipelines. Planned collaborations with researchers from different fields of biology will provide the diversity of use cases needed to surface failure modes, improve skills files and map unmet requirements. These collaborations will focus on accessibility and usability: specifically whether the conversational interface allows biologists to express intent naturally, whether the generated scripts and parameter choices align with domain expectations, and whether errors are communicated in a way that supports correction rather than confusion.

A fundamental challenge in evaluating AI-assisted scientific analysis is that biologists possess tacit intuition about whether an output is plausible. A segmentation that is numerically accurate by pixel overlap metrics may still be biologically incorrect in a way that an expert recognises immediately. This connects to the phenomenon of criteria drift identified by Shankar et al. \cite{validation}: users need pre-existing criteria to grade outputs, yet the grading process itself shapes and revises those criteria, and some criteria emerge only in response to the specific outputs observed. In the context of bioimage analysis, this means that evaluation protocols cannot be fully specified in advance: a biologist asked to assess a tracking result may articulate requirements (minimum track length, handling of cell division events, tolerance for gap-closing) only after seeing cases where the agent's choices diverge from expectation. Addressing this requires iterative validation cycles in which biologists assess outputs and their stated criteria are captured alongside the result, allowing the evaluation standard itself to be treated as a structured output of the collaboration. The state ledger's step-by-step record of agent decisions provides an audit trail that supports this kind of retrospective, criteria-aware validation.

LLM-based agents are inherently stochastic: given the same task, successive runs may produce syntactically distinct scripts that are semantically equivalent, or occasionally diverge in parameter choices or model selection. Quantifying this variability is a prerequisite for scientific use. A repeatability study, running a fixed panel of biological tasks across multiple independent agent sessions and measuring variance in generated code, parameter values, and quantitative outputs, is planned. Metrics of interest include script-level semantic equivalence (whether scripts produce statistically indistinguishable results on the same input), tool-call sequence consistency (whether the agent follows the same reasoning path), and downstream biological concordance (whether cell counts, morphology statistics, or track features agree within acceptable biological variability). Previous work on LLM output consistency provides methodological precedent for this kind of benchmarking \cite{biased_eval}. To support this future assessment, the system already records each run's tool-call sequence, parameter choices, and intermediate outputs in the state ledger, and maintains a versioned script archive with per-attempt failure provenance; these artefacts substantially reduce the cost of systematic repeatability analysis and provide an initial basis for inter-run comparison without requiring additional instrumentation.

The agentic AI ecosystem is developing at a pace that makes any specific implementation a snapshot rather than a final design. New model families with substantially improved reasoning and coding capability are released on timescales of weeks. Open-source model improvements, particularly for code generation and scientific reasoning, are narrowing the gap with proprietary frontier models and may eventually enable fully local deployment without API dependency, which would strengthen the security posture and reduce operating costs for institutions with data governance constraints. The architecture described here is designed to accommodate this evolution: the skills filesystem is model-agnostic, the MCP adapter can expose any compliant host service without changes to agent code, and the conda environment isolation strategy means new computational tools (segmentation models, statistical packages, analysis frameworks) can be added without destabilising existing functionality.

\section{Conclusion}

Agentic-J is a containerised, multi-agent system that makes conversational bioimage analysis available to researchers without programming expertise, while remaining a practical tool for those who have extensive experience already. By keeping the familiar Fiji workspace central, accessible alongside the agent in a single browser tab, the system treats analysis as a collaboration between human expertise and automated assistance. The biologist remains in the loop at every step, exercising judgement where it matters while the agent handles the technical translation from scientific question to executable pipeline.

Two case studies demonstrate the system operating across both of its working modes: a published TrackMate cell-migration analysis \cite{Holme2023} was replicated through the graphical interface, and a multi-step apoptosis read-out for parasite-infected host cells \cite{parasite_paper} was produced as Groovy and Python code by the agent. Quantitative results matched published or expert-derived references, and deviations were flagged by the agent in its own workflow report, leaving an informed user well placed to intervene.

Beyond execution, the system addresses the domain-knowledge gap of general-purpose LLMs in bioimage analysis without retraining the underlying model. A curated knowledge base paired with a plugin-specific skills filesystem, and complemented by recipe and error stores that grow with every session, gives the agent a form of domain competence that scales with use.

The broader practical value of the system will be established through planned collaborations with biologists across disciplines, and through repeatability studies that quantify the consistency of agent-generated pipelines. The architecture is designed to absorb the rapid evolution of the agentic AI landscape: new models, plugins, and external platforms can be integrated without restructuring the core. Agentic-J is a step towards making rigorous, documented, and reproducible bioimage analysis for every biologist who needs it.

\section*{Acknowledgement}

The work of ISAS was supported by the “Ministerium für Kultur und Wissenschaft des Landes Nordrhein-Westfalen” and “Der Regierende Bürgermeister von Berlin, Senatskanzlei Wissenschaft und Forschung.” The work of L.J., Y.Z., J.C. was further supported by the Bundesministerium für Forschung, Technologie und Raumfahrt, BMFTR under the funding reference 161L0272. Xinyi Chen is funded by LESSEN (NWA.1389.20.183) and ELIAS (GA No. 101120237). The work of Institute of Computer Science (University of Tartu) was conducted using the research infrastructure “ELIXIR Estonia” funded by the Estonian Research Council (TARISTU24-TK4).
\noindent

\begin{appendices}
\onecolumn

\section{Supplementary material}\label{supp_mat}
All files related to the biological workflow (mouse stem cell tracking and the apoptotic behaviour of parasite-infested cells) tasks, including input data, scripts, results, and other project files such as costs and phase data can be found at: \url{https://doi.org/10.5281/zenodo.20443685}

\section{Apoptotic behaviour of parasite-infested cell - Extended Material}

\subsection{Task description}\label{task_description_parasite}

\begin{readme}

Background

We work on intracellular parasites that live inside host cells. Sometimes a host cell dies by apoptosis (a controlled form of cell death), and we want to know whether parasites sitting inside such dying cells look different from parasites sitting inside healthy cells. Specifically, we want to compare a fluorescence read-out from the parasites between these two situations.

To get at this, we image the same dish twice. First, while the cells are still alive, we film them under the microscope. Then we take the dish off the microscope, fix the cells (so nothing moves anymore) and stain them with TUNEL, a stain that lights up the nuclei of cells that were undergoing apoptosis. After that we put the dish back on the microscope and image the exact same spot again. So in the end we have a "live" movie and an "after-staining" picture of the same cells.

Our parasites express a fluorescent protein called mKikume. mKikume is special because when you shine UV light on it, it switches from glowing green to glowing red. So each parasite has both a green and a red signal. The number we ultimately care about is the ratio of red to green fluorescence in each individual parasite.

The data files

\texttt{20211028\_Position003.tif} --- The live movie. One position, several time points, 10 minutes between frames.
\begin{itemize}
  \item Channel 1: phase contrast (you can see the host cells in this channel)
  \item Channel 2: mKikume green (parasites, green signal)
  \item Channel 3: mKikume red (parasites, red signal)
\end{itemize}

\texttt{20211028\_TUNEL\_Position003.tif} --- The same spot on the dish, but imaged after the cells were fixed and TUNEL-stained.
\begin{itemize}
  \item Channel 1: phase contrast (same cells as in the live movie, but no longer alive)
  \item Channel 2: TUNEL signal --- bright dots here mean an apoptotic nucleus
  \item Channel 3: DAPI --- labels all nuclei in the field, apoptotic or not
\end{itemize}

Provided ROIs (important change vs the original task)

In this dataset, the parasites and their local background regions were already selected manually.
This means you do NOT need to detect parasites automatically anymore.

Instead, use the provided ROI set: \texttt{Parasite\_and\_background.zip}

ROI naming convention:
\begin{itemize}
  \item Each parasite ROI is named: \texttt{parasite\_\textless n\textgreater}
  \item For each parasite, three nearby background ROIs are provided and named:
    \begin{itemize}
      \item \texttt{background\_for\_parasite\_\textless n\textgreater\_1}
      \item \texttt{background\_for\_parasite\_\textless n\textgreater\_2}
      \item \texttt{background\_for\_parasite\_\textless n\textgreater\_3}
    \end{itemize}
  \item Because multiple ROI batches were merged, some parasite IDs can appear more than once. In that case, the later duplicates are tagged with a suffix: \texttt{\_\_dup2}, \texttt{\_\_dup3}, \ldots{} Example: \texttt{parasite\_1\_\_dup2} has matching backgrounds \texttt{background\_for\_parasite\_1\_1\_\_dup2}, etc.
\end{itemize}

The annoying practical problem (still applies)

Because we had to take the dish off the microscope to fix and stain it, and then put it back, the field of view is not in exactly the same place anymore. When you open the two files side by side, the cells are clearly the same cells, but they sit at slightly different positions in the image, and the whole picture also looks a little tilted compared to the live movie.

So if we just overlay the post-fixation TUNEL channel onto the live movie, the TUNEL dots end up next to the wrong cells, which is obviously useless.

What we actually want is: for every parasite ROI drawn on the last frame of the live movie, we want to know whether the nucleus of its host cell was TUNEL-positive or not. For that to work, the two recordings need to be brought back on top of each other so that the same cell appears in the same place in both pictures.

The phase contrast channel is recorded in both files and shows the same cells in both, so it should be possible to use that to figure out how the two views have to be lined up. Once the two phase contrast views match up, the TUNEL and DAPI channels need to be moved in the same way.

What I want to get out of the analysis (apoptosis read-out)

We use only the LAST frame of the live movie for this part --- everything before it can be ignored.

For every parasite ROI in the provided ROI set, I want to know two things:
\begin{enumerate}[label=(\alph*)]
  \item Is the host cell that this parasite sits in apoptotic? In other words, does its nucleus light up in the TUNEL channel of the (aligned) post-fix image? The DAPI channel can help to locate nuclei in general; TUNEL-positive means it is also bright in the TUNEL channel, not just in DAPI.
  \item What is the corrected red-to-green ratio of that parasite?
\end{enumerate}

By "corrected" I mean: the parasite itself has some red and green signal, but the surrounding area also has baseline fluorescence that should not be counted as parasite signal.

Because background ROIs are already provided for each parasite, the simplest correction is:
\begin{itemize}
  \item measure the mean intensity of the parasite ROI in green and red
  \item measure the mean intensity in each of the three background ROIs (green and red)
  \item average the three backgrounds per channel
  \item subtract that average background from the parasite mean in each channel
\end{itemize}

After background subtraction, compute per parasite:
\begin{quote}
$R/G$ = (background-corrected red) / (background-corrected green)\\
final value = $10 - R/G$
\end{quote}

The "$10 - R/G$" is just the convention we use in our group, please keep it.

What I'd ultimately like to see, per parasite:
\begin{itemize}
  \item which parasite it is (ROI name/ID, ideally with a marked-up image so I can check),
  \item was its host cell TUNEL-positive (yes/no),
  \item its final $10 - R/G$ value.
\end{itemize}

A simple table of these numbers, with the TUNEL+ and TUNEL- parasites clearly separable, is what we then use for the statistics in the paper.

\end{readme}

\subsection{Workflow documentation}\label{workflow_doc_parasite}

\begin{readme}

{\large\bfseries Workflow Documentation}\par
\medskip
\textbf{Project name:} parasite\_apoptosis\_roi\_alignment\\
\textbf{Date:} 2026-05-22\\
\textbf{Workflow type:} New Workflow (example / use-case execution)

\medskip\hrule\medskip

\noindent\textbf{1. Scientific Goal}\par\smallskip
Quantify parasite reporter conversion (mKikume red/green ratio) on the \textbf{last frame of a live movie}, and determine whether each parasite is in a \textbf{TUNEL-positive vs TUNEL-negative host cell nucleus} by aligning a post-fix TUNEL/DAPI image to the live phase-contrast image.

\medskip\hrule\medskip

\noindent\textbf{2. Software Components \& Versions}\par\smallskip
(From \texttt{logs/environment\_metadata.log})

\smallskip
\begin{tabularx}{\linewidth}{@{} l r >{\raggedright\arraybackslash}X @{}}
\toprule
\textbf{Component} & \textbf{Version} & \textbf{Role} \\
\midrule
ImageJ/Fiji & 2.16.0/1.54p & Alignment, ROI-based measurements, nuclei segmentation, TUNEL calling \\
Python      & 3.13.13      & Statistics + plotting \\
pandas      & 3.0.2        & Data wrangling \\
numpy       & 2.4.3        & Numeric computation \\
scipy       & 1.17.1       & Statistical tests \\
matplotlib  & 3.10.9       & Plot rendering \\
seaborn     & 0.13.2       & Plot styling \\
\bottomrule
\end{tabularx}

\smallskip
Recommended Fiji plugin used for registration: \textbf{TurboReg / StackReg}.

\medskip\hrule\medskip

\noindent\textbf{3. Processing Sequence}\par\smallskip
\begin{enumerate}
  \item \textbf{IO check (memory-safe)} (\texttt{phase4a\_io\_check.groovy})
    \begin{itemize}
      \item Verified that the live movie TIFF, post-fix TIFF, and ROI zip load correctly.
      \item Opened large TIFFs as \textbf{virtual stacks} to avoid Java heap OutOfMemory.
      \item Output: \texttt{data/IO\_check\_summary.csv}
    \end{itemize}

  \item \textbf{Rigid registration of post-fix $\rightarrow$ live (TurboReg on phase)} (\texttt{phase4b\_alignment\_turboreg\_single\_verification.groovy})
    \begin{itemize}
      \item Extracted the \textbf{last frame} of the live movie.
      \item Used \textbf{phase contrast channel} as the registration reference.
      \item Estimated a rigid-body transform (TurboReg) to align post-fix phase to live phase.
      \item Applied the same transform to post-fix \textbf{TUNEL} and \textbf{DAPI} channels.
      \item Saved QC overlay and aligned channel images.
      \item Manual verification: user confirmed alignment looked good.
    \end{itemize}

  \item \textbf{ROI-based parasite intensity measurement on live last frame} (\texttt{phase4b\_roi\_intensity\_measurement\_live\_lastframe.groovy})
    \begin{itemize}
      \item Loaded ROI set from \texttt{Parasite\_and\_background.zip}.
      \item For each \texttt{parasite\_*} ROI, measured mean intensities in:
        \begin{itemize}
          \item live \textbf{green} channel (mKikume green)
          \item live \textbf{red} channel (mKikume red)
        \end{itemize}
      \item Background correction:
        \begin{itemize}
          \item matched background ROIs \texttt{background\_for\_parasite\_\textless n\textgreater\_[1-3]}
          \item averaged the 3 backgrounds and subtracted from parasite mean
          \item handled duplicate parasite ROIs (\texttt{\_\_dup} suffix) via base-ID fallback.
        \end{itemize}
      \item Computed ratios:
        \begin{itemize}
          \item \texttt{R\_over\_G}
          \item \texttt{final\_10\_minus\_R\_over\_G = 10 - (R/G)}
        \end{itemize}
      \item Outputs: \texttt{data/parasite\_ratio\_live\_lastframe.csv}, \texttt{processed\_images/parasite\_rois\_on\_live\_lastframe.tif}
    \end{itemize}

  \item \textbf{TUNEL+/$-$ calling from aligned post-fix image} (\texttt{phase4b\_tunel\_classification\_postfix\_aligned.groovy})
    \begin{itemize}
      \item Segmented nuclei from \textbf{aligned DAPI}.
      \item Measured mean TUNEL intensity per nucleus.
      \item Classified nuclei as TUNEL+/$-$ using \textbf{Otsu threshold} on nucleus mean TUNEL values (threshold reported as $\sim$227.6 in the run log/ledger).
      \item Assigned each parasite ROI to a nucleus:
        \begin{itemize}
          \item first by ROI containment
          \item else by nearest nucleus within a maximum distance (if none, parasite marked unassigned).
        \end{itemize}
      \item Outputs:
        \begin{itemize}
          \item \texttt{data/nuclei\_tunel\_measurements.csv}
          \item \texttt{data/tunel\_calls\_per\_parasite.csv}
          \item \texttt{processed\_images/labeled\_nuclei.tif}
          \item \texttt{processed\_images/tunel\_positive\_overlay.tif}
        \end{itemize}
    \end{itemize}

  \item \textbf{Merge results into one per-parasite table} (\texttt{phase4b\_merge\_results\_table.groovy})
    \begin{itemize}
      \item Joined live ratio table with parasite-level TUNEL calls.
      \item Output: \texttt{data/Parasite\_Apoptosis\_Final\_Table.csv}
    \end{itemize}

  \item \textbf{Statistics (assigned-only)} (\texttt{phase4c\_statistics\_final\_10minus\_RoverG\_by\_TUNEL.py})
    \begin{itemize}
      \item Compared \texttt{final\_10\_minus\_R\_over\_G} between TUNEL+ and TUNEL$-$ groups.
      \item Reported:
        \begin{itemize}
          \item Mann--Whitney U $p \approx 0.0161$
          \item Cliff's delta $\approx 0.661$
          \item Welch t-test $p \approx 0.000892$
        \end{itemize}
      \item Outputs: \texttt{data/Statistics\_Results.csv}, \texttt{data/Statistics\_Summary.md}
    \end{itemize}

  \item \textbf{Plotting (assigned-only)} (\texttt{phase4d\_plotting\_optional\_final10minusRoverG\_by\_TUNEL.py})
    \begin{itemize}
      \item Generated:
        \begin{itemize}
          \item violin/box/strip plot of \texttt{final\_10\_minus\_R\_over\_G} by TUNEL status (annotated with Mann--Whitney p)
          \item count plot by TUNEL status
        \end{itemize}
      \item Outputs (PNG + SVG): see \texttt{figures/}.
    \end{itemize}
\end{enumerate}

\medskip\hrule\medskip

\noindent\textbf{4. Key Settings \& Parameters}\par\smallskip

\noindent\resizebox{\linewidth}{!}{%
\begin{tabular}{@{} l l l p{5.5cm} @{}}
\toprule
\textbf{Parameter} & \textbf{Value} & \textbf{Script} & \textbf{Rationale} \\
\midrule
Live frame used &
last frame (T=263) &
\texttt{phase4b\_*} &
Workflow goal specifies last live frame \\

Registration &
TurboReg rigid-body on phase &
\texttt{phase4b\_alignment\_turboreg\_single\_verification.groovy} &
Phase channel shared pre/post fixation; rigid alignment sufficient for same field \\

Background per parasite &
average of 3 ROI backgrounds &
\texttt{phase4b\_roi\_intensity\_measurement\_live\_lastframe.groovy} &
Reduces noise vs single background ROI \\

Duplicate parasite ROI handling &
exact match, else base-ID fallback for \texttt{\_\_dup} &
\texttt{phase4b\_roi\_intensity\_measurement\_live\_lastframe.groovy} &
Ensures duplicates inherit the intended backgrounds \\

Nucleus segmentation &
DAPI-based mask &
\texttt{phase4b\_tunel\_classification\_postfix\_aligned.groovy} &
DAPI provides nuclear boundary \\

TUNEL thresholding &
Otsu on nucleus mean TUNEL &
\texttt{phase4b\_tunel\_classification\_postfix\_aligned.groovy} &
Objective thresholding for binary call \\

Significance level &
$p < 0.05$ &
Python stats script &
Conventional threshold \\
\bottomrule
\end{tabular}%
}

\medskip\hrule\medskip

\noindent\textbf{5. Input Data}\par\smallskip
\begin{itemize}
  \item \textbf{Live movie:} \texttt{/app/data/tasks/Parasite\_use\_case/20211028\_Position003.tif}
    \begin{itemize}
      \item 3 channels: phase, mKikume green, mKikume red
      \item time series; last frame used for measurements
    \end{itemize}
  \item \textbf{Post-fix:} \texttt{/app/data/tasks/Parasite\_use\_case/20211028\_TUNEL\_Position003.tif}
    \begin{itemize}
      \item 3 channels: phase, TUNEL, DAPI
    \end{itemize}
  \item \textbf{ROI set:} \texttt{/app/data/tasks/Parasite\_use\_case/Parasite\_and\_background.zip}
    \begin{itemize}
      \item \texttt{parasite\_\textless n\textgreater} plus \texttt{background\_for\_parasite\_\textless n\textgreater\_[1-3]} (with possible \texttt{\_\_dup} suffixes)
    \end{itemize}
\end{itemize}

Image metadata (from extraction/ledger): 16-bit; pixel size $\approx 0.322\,\upmu$m/px.

\medskip\hrule\medskip

\noindent\textbf{6. Output Data}\par\smallskip

\noindent\textit{Primary results}
\begin{itemize}
  \item \textbf{Final combined table:} \texttt{data/Parasite\_Apoptosis\_Final\_Table.csv}
\end{itemize}

\smallskip\noindent\textit{Intermediate tables}
\begin{itemize}
  \item \texttt{data/parasite\_ratio\_live\_lastframe.csv}
  \item \texttt{data/tunel\_calls\_per\_parasite.csv}
  \item \texttt{data/nuclei\_tunel\_measurements.csv}
  \item \texttt{data/alignment\_report.csv}
\end{itemize}

\smallskip\noindent\textit{QC images}
\begin{itemize}
  \item \texttt{processed\_images/alignment\_QC\_overlay.tif}
  \item \texttt{processed\_images/parasite\_rois\_on\_live\_lastframe.tif}
  \item \texttt{processed\_images/labeled\_nuclei.tif}
  \item \texttt{processed\_images/tunel\_positive\_overlay.tif}
\end{itemize}

\smallskip\noindent\textit{Statistics + figures}
\begin{itemize}
  \item Statistics: \texttt{data/Statistics\_Results.csv}, \texttt{data/Statistics\_Summary.md}
  \item Figures (300 DPI PNG + SVG):
    \begin{itemize}
      \item \texttt{figures/Fig1\_final10minusRoverG\_by\_TUNEL\_assigned\_only.(png\textbar svg)}
      \item \texttt{figures/Fig2\_counts\_by\_TUNEL\_assigned\_only.(png\textbar svg)}
    \end{itemize}
\end{itemize}

\medskip\hrule\medskip

\noindent\textbf{7. Statistical Analysis}\par\smallskip
\begin{itemize}
  \item \textbf{Groups:} parasites assigned to \textbf{TUNEL+} vs \textbf{TUNEL$-$} nuclei (assigned-only)
  \item \textbf{Outcome:} \texttt{final\_10\_minus\_R\_over\_G}
  \item \textbf{Tests used:} Mann--Whitney U; Welch's t-test; effect size Cliff's delta
  \item \textbf{Significance threshold:} 0.05
  \item \textbf{Outliers:} not explicitly removed (analysis is on the assigned-only subset)
\end{itemize}

\medskip\hrule\medskip

\noindent\textbf{8. Manual Steps (ROI / User Verification)}\par\smallskip
\begin{itemize}
  \item User visually verified the \textbf{registration QC overlay} after TurboReg alignment (``looks good'').
  \item Parasite and background ROIs were provided as an input ROI zip (no manual drawing required during this run).
\end{itemize}

\medskip\hrule\medskip

\noindent\textbf{9. Rationale}\par\smallskip
\begin{itemize}
  \item \textbf{Use phase channel for alignment}: phase contrast is present in both live and post-fix acquisitions.
  \item \textbf{Apply the same transform to fluorescence channels}: preserves relative channel geometry after alignment.
  \item \textbf{ROI-based parasite measurement}: the use-case provides curated parasite ROIs; avoids segmentation complexity.
  \item \textbf{DAPI-based nuclei segmentation}: creates nuclear regions for consistent TUNEL quantification.
\end{itemize}

\medskip\hrule\medskip

\noindent\textbf{10. Limitations \& Known Issues}\par\smallskip
\begin{itemize}
  \item Some parasite ROIs may be \textbf{unassigned} to any nucleus (reported in the final table \texttt{assignment\_qc}). These should be reviewed/excluded if the biological question requires a nucleus call.
  \item Thresholding-based TUNEL calls depend on staining/background; Otsu is objective but may not match expert calls in all datasets.
  \item Java heap constraints required memory-safe (virtual-stack) IO; avoid opening full time-series as non-virtual in Fiji.
\end{itemize}

\medskip\hrule\medskip

\noindent\textbf{11. Reproducibility}\par\smallskip
\begin{itemize}
  \item \textbf{Groovy scripts:} \texttt{scripts/imagej/*.groovy}
  \item \textbf{Python scripts:} \texttt{scripts/imagej/*.py} and \texttt{scripts/python/*.py}
  \item \textbf{All outputs:} within this project root.
  \item \textbf{Example data:} \texttt{/app/data/tasks/Parasite\_use\_case/}
  \item \textbf{Public access / DOI:} [TO BE FILLED if you plan to publish the workflow]
\end{itemize}

\end{readme}

\subsection{QA report}\label{QA_report_parasite}

\begin{readme}

{\large\bfseries QA Checklist Report}\par
\medskip
\textbf{Project:} parasite\_apoptosis\_roi\_alignment\\
\textbf{Date:} 2026-05-22\\
\textbf{Workflow type:} New Workflow

\smallskip
\noindent\textbf{Overall status:}
\begin{itemize}
  \item Workflow: \textbf{5/6 Minimal passed} \textbar{} \textbf{3/4 Recommended passed}
  \item Image Publishing: \textbf{4/12 Minimal passed} \textbar{} \textbf{0/2 Recommended passed}
\end{itemize}

\medskip\hrule\medskip

\noindent{\bfseries PART A: WORKFLOW STANDARDS}\par

\smallskip
\noindent\textit{MINIMAL Requirements}\par\smallskip

\noindent\resizebox{\linewidth}{!}{%
\begin{tabular}{@{} p{3.3cm} p{1.8cm} p{8cm} @{}}
\toprule
\textbf{Item} & \textbf{Status} & \textbf{Evidence} \\
\midrule
Cite components and platform &
\textcolor{green!50!black}{\textbf{PASS}} &
\texttt{Workflow\_Documentation.md} + \texttt{logs/environment\_metadata.log} list Fiji/ImageJ and Python libs with versions (e.g., Fiji 2.16.0/1.54p; Python 3.13.13; pandas/numpy/scipy/matplotlib/seaborn). \\

Describe sequence &
\textcolor{green!50!black}{\textbf{PASS}} &
\texttt{Workflow\_Documentation.md} Section 3 lists IO check $\rightarrow$ TurboReg alignment $\rightarrow$ ROI measurement $\rightarrow$ TUNEL classification $\rightarrow$ merge $\rightarrow$ stats $\rightarrow$ plotting. \\

Key settings &
\textcolor{green!50!black}{\textbf{PASS}} &
Key parameters documented in \texttt{Workflow\_Documentation.md} and script docs: TurboReg rigidBody; nuclei segmentation sigma=1.0, Otsu, size $\geq$ 80\,px; maxAssignDistance=50\,px; ratio definition $10-(R/G)$. \\

Example data and code &
\textcolor{orange!80!black}{\textbf{PARTIAL}} &
Scripts are present under \texttt{scripts/}, but \texttt{raw\_images/} is empty; example inputs are referenced as \texttt{/app/data/tasks/Parasite\_use\_case/...} rather than copied into project. \\

Manual ROI &
\textcolor{green!50!black}{\textbf{PASS}} &
Manual/interactive step documented: user verified alignment QC overlay (\texttt{Workflow\_Documentation.md} Section 8). ROI drawing itself not performed in-run; ROIs provided as input zip. \\

Exact versions &
\textcolor{green!50!black}{\textbf{PASS}} &
Exact versions recorded in \texttt{logs/environment\_metadata.log} and repeated in \texttt{Workflow\_Documentation.md}. \\
\bottomrule
\end{tabular}%
}

\smallskip
\noindent\textit{RECOMMENDED Requirements}\par\smallskip

\noindent\resizebox{\linewidth}{!}{%
\begin{tabular}{@{} p{3.3cm} p{1.8cm} p{8cm} @{}}
\toprule
\textbf{Item} & \textbf{Status} & \textbf{Evidence} \\
\midrule
All settings documented &
\textcolor{orange!80!black}{\textbf{PARTIAL}} &
Many parameters are explicit (e.g., Gaussian sigma, min nucleus size, max assign distance, TurboReg landmarks), but not all image display/export settings are documented (e.g., LUTs/B\&C, calibration retention checks, annotation sizes). \\

Public example data and code &
\textcolor{red!70!black}{\textbf{FAIL}} &
\texttt{Workflow\_Documentation.md} has ``Public access / DOI: [TO BE FILLED]''; no URL/DOI/repository provided. \\

Rationale &
\textcolor{green!50!black}{\textbf{PASS}} &
Rationale section present (\texttt{Workflow\_Documentation.md} Section 9) explaining phase-based alignment, reuse transform, ROI-based measurement, DAPI segmentation. \\

Limitations &
\textcolor{green!50!black}{\textbf{PASS}} &
Limitations listed (\texttt{Workflow\_Documentation.md} Section 10) including unassigned parasites, Otsu dependence, Fiji memory constraints. \\
\bottomrule
\end{tabular}%
}

\smallskip
\noindent\textit{IDEAL Requirements}\par\smallskip

\noindent\resizebox{\linewidth}{!}{%
\begin{tabular}{@{} p{3.3cm} p{1.8cm} p{8cm} @{}}
\toprule
\textbf{Item} & \textbf{Status} & \textbf{Evidence} \\
\midrule
Screen recording or tutorial &
\textcolor{red!70!black}{\textbf{FAIL}} &
No tutorial/screencast link/file found in project tree. \\

Easy install / container &
\textcolor{red!70!black}{\textbf{FAIL}} &
No \texttt{requirements.txt}, \texttt{environment.yml}, or Dockerfile found; only \texttt{logs/environment\_metadata.log} lists versions. \\
\bottomrule
\end{tabular}%
}

\medskip\hrule\medskip

\noindent{\bfseries PART B: IMAGE PUBLISHING STANDARDS}\par

\smallskip
\noindent\textit{Image Format --- MINIMAL}\par\smallskip

\noindent\resizebox{\linewidth}{!}{%
\begin{tabular}{@{} p{3.3cm} p{1.8cm} p{8cm} @{}}
\toprule
\textbf{Item} & \textbf{Status} & \textbf{Evidence} \\
\midrule
Focus on relevant content &
\textcolor{orange!80!black}{\textbf{PARTIAL}} &
QC images are produced (ROI overlay; alignment overlay), but no explicit crop/rotate/resize to highlight relevant regions is documented in scripts. \\

Separate individual images &
\textcolor{orange!80!black}{\textbf{PARTIAL}} &
Multiple individual TIFF QC outputs exist in \texttt{processed\_images/} (aligned channels, overlays), but no explicit export set for per-channel presentation beyond these QC outputs; \texttt{processed\_images/channels/} is empty. \\

Show example image &
\textcolor{red!70!black}{\textbf{FAIL}} &
\texttt{raw\_images/} is empty; example inputs exist only as external paths \texttt{/app/data/tasks/Parasite\_use\_case/...} (referenced in docs/scripts). \\
\bottomrule
\end{tabular}%
}

\smallskip
\noindent\textit{Image Colors \& Channels --- MINIMAL}\par\smallskip

\noindent\resizebox{\linewidth}{!}{%
\begin{tabular}{@{} p{3.3cm} p{1.8cm} p{8cm} @{}}
\toprule
\textbf{Item} & \textbf{Status} & \textbf{Evidence} \\
\midrule
Annotation of channels visible &
\textcolor{green!50!black}{\textbf{PASS}} &
Channel roles described in ledger + \texttt{Workflow\_Documentation.md} (phase used for registration; TUNEL and DAPI; live green/red for mKikume). \\

Report B\&C adjustments &
\textcolor{red!70!black}{\textbf{FAIL}} &
No brightness/contrast adjustment parameters or statements found in Groovy/Python scripts or docs; QC overlays are saved without reporting display ranges. \\

Grayscale for each channel &
\textcolor{orange!80!black}{\textbf{PARTIAL}} &
Separate aligned TUNEL and DAPI TIFFs are saved (\texttt{processed\_images/aligned\_postfix\_TUNEL.tif}, \texttt{aligned\_postfix\_DAPI.tif}), but no documented saving of individual live channels or a structured \texttt{processed\_images/channels/} export (folder exists but empty). \\

Same adjustments for comparisons &
\textcolor{red!70!black}{\textbf{FAIL}} &
Alignment QC overlay created via \texttt{Merge Channels} but no explicit consistent display scaling/B\&C synchronization is documented. \\

Color-blind accessible &
\textcolor{green!50!black}{\textbf{PASS}} &
Plot scripts use explicit colorblind-friendly palette (\texttt{\#D55E00} and gray) and save SVG/PNG at 300\,DPI (\texttt{phase4d\_plotting\_assigned\_only.py} and \texttt{phase4d\_plotting\_optional\_final10minusRoverG\_by\_TUNEL.py}). \\
\bottomrule
\end{tabular}%
}

\smallskip
\noindent\textit{Image Colors \& Channels --- RECOMMENDED}\par\smallskip

\noindent\resizebox{\linewidth}{!}{%
\begin{tabular}{@{} p{3.3cm} p{1.8cm} p{8cm} @{}}
\toprule
\textbf{Item} & \textbf{Status} & \textbf{Evidence} \\
\midrule
Provide intensity scales (scale bars) &
\textcolor{red!70!black}{\textbf{FAIL}} &
No intensity calibration bars are created; no scripts call an intensity scale function or annotate LUT scales. \\
\bottomrule
\end{tabular}%
}

\smallskip
\noindent\textit{Image Annotation --- MINIMAL}\par\smallskip

\noindent\resizebox{\linewidth}{!}{%
\begin{tabular}{@{} p{3.3cm} p{1.8cm} p{8cm} @{}}
\toprule
\textbf{Item} & \textbf{Status} & \textbf{Evidence} \\
\midrule
Add scale information &
\textcolor{red!70!black}{\textbf{FAIL}} &
No scale bar added in scripts (no \texttt{IJ.run("Scale Bar...")}); processed TIFFs appear to be raw QC outputs without scale annotation. \\

Explain all annotations &
\textcolor{orange!80!black}{\textbf{PARTIAL}} &
QC overlays exist (ROI labels; nucleus +/$-$ labels), but annotation meaning is not fully documented for publication (e.g., cyan vs red nuclei in \texttt{tunel\_positive\_overlay.tif}). \\

Annotations legible &
\textcolor{orange!80!black}{\textbf{PARTIAL}} &
ROI and nucleus labels are added (TextRoi) but font size/line width standards are not fully specified; nucleus stroke width set to 1.5 ($<$2 recommended) in \texttt{phase4b\_tunel\_classification\_postfix\_aligned.groovy}. \\

Annotations don't obscure data &
\textcolor{orange!80!black}{\textbf{PARTIAL}} &
Labels are placed at ROI bounding-box top-left (could obscure content); no documented strategy to avoid obscuring key data. \\
\bottomrule
\end{tabular}%
}

\smallskip
\noindent\textit{Image Availability --- MINIMAL}\par\smallskip

\noindent\resizebox{\linewidth}{!}{%
\begin{tabular}{@{} p{3.3cm} p{1.8cm} p{8cm} @{}}
\toprule
\textbf{Item} & \textbf{Status} & \textbf{Evidence} \\
\midrule
Images shared (lossless compression) &
\textcolor{green!50!black}{\textbf{PASS}} &
Processed outputs are saved as TIFF (\texttt{processed\_images/*.tif}), not JPEG. \\
\bottomrule
\end{tabular}%
}

\medskip\hrule\medskip

\noindent\textbf{Action Items}\par\smallskip

\noindent\textit{WORKFLOW --- Critical Failures}\par
\noindent List every \textcolor{red!70!black}{\textbf{FAIL}} from workflow MINIMAL requirements:
\begin{itemize}
  \item Example data and code: add at least one representative raw input image to \texttt{raw\_images/} (or a small cropped sample) and/or a note explaining where the example data can be retrieved.
\end{itemize}

\smallskip
\noindent\textit{WORKFLOW --- Recommended Improvements}\par
\noindent List every \textcolor{red!70!black}{\textbf{FAIL}} or \textcolor{orange!80!black}{\textbf{PARTIAL}} from workflow RECOMMENDED requirements:
\begin{itemize}
  \item All settings documented: add explicit documentation of remaining parameters affecting outputs (e.g., any LUT/B\&C choices, overlay annotation settings, calibration preservation checks).
  \item Public example data and code: fill in the \texttt{[TO BE FILLED]} DOI/URL/repository location in \texttt{Workflow\_Documentation.md}.
\end{itemize}

\smallskip
\noindent\textit{IMAGE PUBLISHING --- Critical Failures}\par
\noindent List every \textcolor{red!70!black}{\textbf{FAIL}} from image publishing MINIMAL requirements:
\begin{itemize}
  \item Show example image: place at least one example raw image in \texttt{raw\_images/} (or document an accessible location and include a small representative crop).
  \item Report B\&C adjustments: document any brightness/contrast changes (min/max display values or statement of ``linear, identical scaling across conditions'') for all published images.
  \item Same adjustments for comparisons: ensure comparison images (e.g., aligned vs live phase overlay) use identical display scaling and document it.
  \item Add scale information: add scale bars (and confirm pixel size calibration is correct) for publication images.
\end{itemize}

\smallskip
\noindent\textit{IMAGE PUBLISHING --- Recommended Improvements}\par
\noindent List every \textcolor{red!70!black}{\textbf{FAIL}} or \textcolor{orange!80!black}{\textbf{PARTIAL}} from image publishing RECOMMENDED requirements:
\begin{itemize}
  \item Provide intensity scales (scale bars): add intensity/LUT scale bars where intensity comparisons are visually interpreted.
  \item Focus on relevant content: add cropping/inset strategy and document any resizing/rotation.
  \item Separate individual images: export a clean set of per-channel grayscale images to \texttt{processed\_images/channels/} (and a merged figure if needed).
  \item Explain all annotations / legibility / obscuration: document annotation conventions (colors, line widths $\geq$2, font size $\geq$12) and ensure labels do not cover key structures.
\end{itemize}

\end{readme}


\begin{figure*}[htpb]
    \centering
    \includegraphics[width=\linewidth]{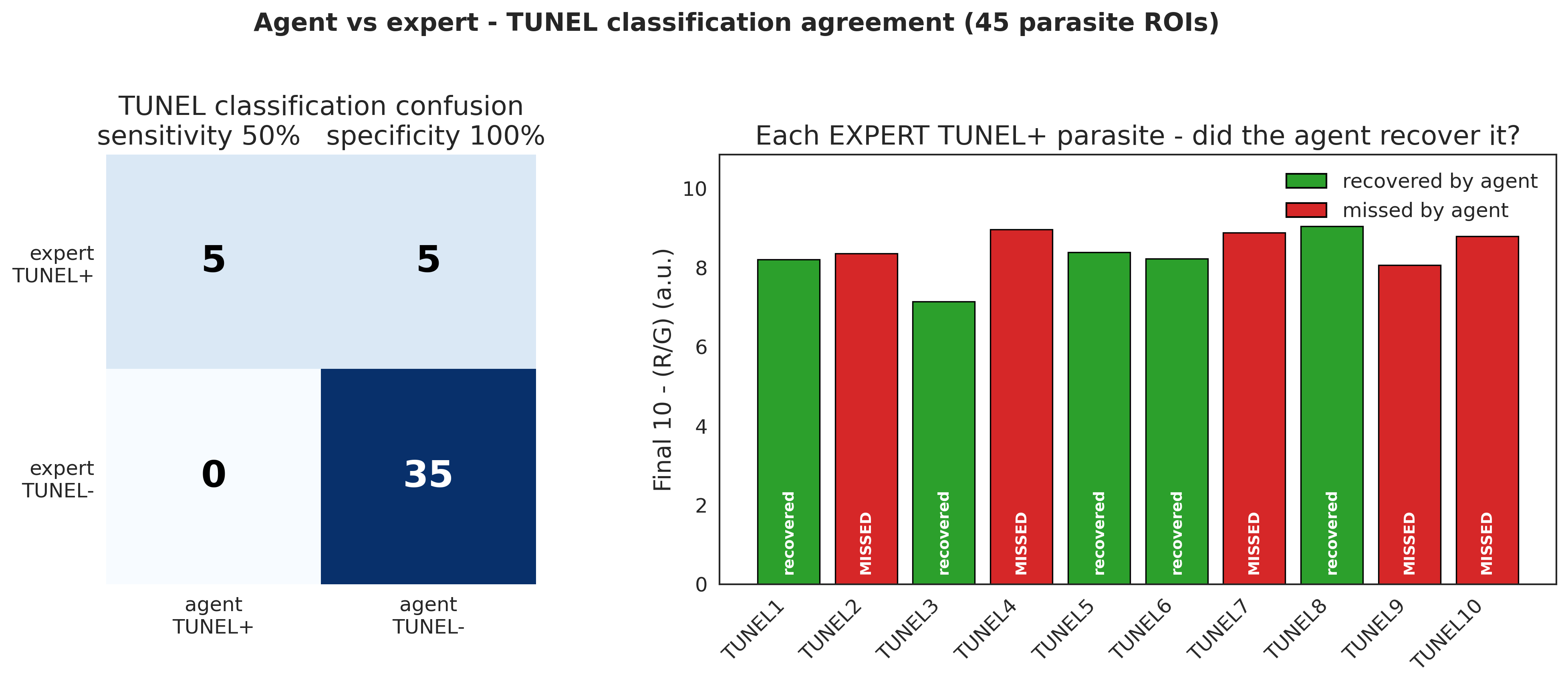}
    \caption{Agreement between the expert ground-truth and the agent-derived TUNEL
classifications across all 45 parasite ROIs. Left: the confusion matrix; the
agent produced no false positives (specificity 100\%) but recovered only half
of the expert TUNEL-positive parasites (sensitivity 50\%). Right: a per-parasite
breakdown of the ten expert TUNEL-positive parasites, showing which were
recovered (green) and which were missed (red) by the agent. The missed parasites
are largely those that could not be assigned to a nearby nucleus following image
registration and nucleus assignment.}
\label{fig:classification_agreement}

\end{figure*}

\section{RAG knowledge base inventory}\label{app:rag-inventory}

The \texttt{BioimageAnalysisDocs} Qdrant collection referenced in
Section~3.3.1 was populated, at container build time, with the
open-source resources listed below. Each item is chunked into hybrid
sparse/dense embeddings; all sources are redistributable under
CC-BY-4.0, BSD-3, or an equivalent permissive licence.

\paragraph{Biomedical image-processing textbooks.}
\begin{itemize}
  \item Bankhead, \emph{Introduction to Bioimage
        Analysis}~\cite{Bankhead2023}.
  \item Bankhead, \emph{Analyzing fluorescence microscopy images with
        ImageJ} (Queen's University Belfast,
        2014)~\cite{bankhead2014analyzing}.
  \item Miura \& Sladoje (eds.), \emph{Bioimage Data Analysis Workflows
        -- Advanced Components and Methods}, Springer
        2022~\cite{miura2022bioimage}.
\end{itemize}

\paragraph{Community best-practice guides.}
\begin{itemize}
  \item Senft et al., \emph{A biologist's guide to planning and
        performing quantitative bioimaging
        experiments}~\cite{senft2023biologist}, together with its
        companion \emph{MicroscopyForBeginnersReferenceGuide} /
        bioimagingguide.org website~\cite{cobanih2023microscopyguide}.
\end{itemize}

\paragraph{ImageJ training material and reference manuals.}
\begin{itemize}
  \item Ferreira \& Rasband, \emph{ImageJ User Guide -- IJ
        1.46r}~\cite{ferreira2012imagej}.
  \item Nowell, \emph{Fiji Training Manual (v6.5)}, Monash University
        2023~\cite{nowell2023fiji}.
  \item NEUBIAS Community, \emph{Bioimage Analysis Training
        Resources}~\cite{neubias2024trainingresources}.
\end{itemize}

\paragraph{Plugin-specific documentation.}
\begin{itemize}
  \item MorphoLibJ -- canonical reference paper~\cite{legland2016morpholibj}
        and NEUBIAS Academy@Home webinar
        slides~\cite{legland2020morpholibj_slides}.
  \item 3D ImageJ Suite -- NEUBIAS Academy@Home webinar
        slides by Boudier~\cite{boudier2020_3dsuite_slides};
        primary citation Ollion et al.~\cite{ollion2013tango}.
\end{itemize}

\paragraph{Jupyter notebook collections.}
\begin{itemize}
  \item Haase et al., \emph{BioImageAnalysisNotebooks}, Zenodo
        \mbox{v.~gpt-2024.1.6}~\cite{haase2024bioimage}.
\end{itemize}

\end{appendices}
\bibliography{sn-bibliography}

\end{document}